\documentclass[twocolumn,preprintnumbers,amsmath,amssymb,superscriptaddress]{revtex4}

\usepackage{graphicx}
\usepackage{dcolumn}
\usepackage{bm}
\usepackage{soul}
\usepackage{color}
\usepackage{epstopdf}
\usepackage[version=3]{mhchem}
\usepackage{lipsum}
\usepackage[outercaption]{sidecap}
\usepackage{floatrow}
\begin{document}

\title{Molecular relaxations in supercooled liquid and glassy states of amorphous gambogic acid: dielectric spectroscopy, calorimetry and theoretical approach}

\author{Anh D. Phan}
\affiliation{Faculty of Materials Science and Engineering, Phenikaa Institute for Advanced Study and Faculty of Information Technology, Artificial Intelligence Laboratory, Phenikaa University, Hanoi 12116, Vietnam}
\email{anh.phanduc@phenikaa-uni.edu.vn}
\affiliation{Department of Nanotechnology for Sustainable Energy, School of Science and Technology, Kwansei Gakuin University, Sanda, Hyogo 669-1337, Japan}

\author{Tran Thi Thu Thuy}
\affiliation{Institute of Natural Products Chemistry, Vietnam Academy of Science and Technology, 18 Hoang Quoc Viet, Hanoi 12116, Vietnam}
\email{thuy.tran@inpc.vast.vn}
\author{Nguyen Thi Kim An}
\affiliation{Faculty of Chemical Technology, Hanoi University of Industry, 298 Cau Dien street, North Tu Liem district, Hanoi 12116, Vietnam}
\affiliation{Graduate University of science and Technology, Vietnam Academy of Science and Technology, 18 Hoang Quoc Viet, Hanoi 12116, Vietnam}
\author{Justyna Knapik-Kowalczuk}
\affiliation{Institute of Physics, University of Silesia, SMCEBI, 75 Pulku Piechoty 1a, 41-500 Chorzow, Poland}
\author{Marian Paluch}
\affiliation{Institute of Physics, University of Silesia, SMCEBI, 75 Pulku Piechoty 1a, 41-500 Chorzow, Poland}
\author{Katsunori Wakabayashi}
\affiliation{Department of Nanotechnology for Sustainable Energy, School of Science and Technology, Kwansei Gakuin University, Sanda, Hyogo 669-1337, Japan}
\date{\today}

\begin{abstract}
The relaxation dynamics and thermodynamic properties of supercooled and glassy gambogic acid are investigated using both theory and experiment. We measure the temperature dependence of the relaxation times in three polymorphs ($\alpha-$, $\beta-$, and $\gamma-$ form). To gain insight into the relaxation processes, we propose a theoretical approach to quantitatively understand nature of these three relaxations. The $\alpha-$relaxation captures cooperative motions of molecules while the $\beta-$process is mainly governed by local dynamics of a single molecule within the cage formed by its nearest neighbors. Based on quantitative agreement between theory and experimental data, our calculations clearly indicate that the $\beta-$process is a precursor of the structural relaxation and intramolecular motions are responsible for the $\gamma-$relaxation. Moreover, the approach is exploited to study effects of the heating process on alpha relaxation. We find that the heating rate varies logarithmically with $T_g$ and $1000/T_g$. These variations are qualitatively consistent with many prior studies.
\end{abstract}

\maketitle

\section{Introduction}
Although the development of science and nanotechnology has improved human daily life to be better and more comfortable, it has led to various consequences including serious diseases. According to the World Health Organization \cite{23}, approximately 68 $\%$ of the 56.4 million deaths worldwide in 2015 were due to diseases. Many cases such as cancer, diabetes and HIV cannot be cured by drugs. Additionally, poor water-solubility issues are challenging the pharmaceutical industry and significantly reduce the efficiency of medical treatment \cite{24,25}. Manufacturing and storage conditions also directly affect pharmaceutical quality and stability. Thus, it is necessary to intensively investigate the properties of medical drugs.

Among the different kinds of medical drugs, scientists and pharmaceutical industries have focused on amorphous drugs \cite{24,25,26,27,28} since the enhancement of solubility and bioavailability compares to the crystalline replicas. The amorphous form of a drug is obtained by cooling a pharmaceutical liquid at a fast rate to avoid crystallization. The molecular mobility in an amorphous material is characterized by the structural/alpha relaxation time in the supercooled-liquid and glassy states. The material has a long-range-disordered structure, which is temperature dependent. Its physical behaviors are liquid-like at high temperatures. However, at low temperatures, structural relaxation process is significantly slowed down. Physical mechanisms underlying the glassy dynamics remain mysterious.

Many methods can be used to investigate the glassy dynamics of amorphous materials \cite{24,25}. The most popular technique is differential scanning calorimetry (DSC). This technique characterizes several types of thermal behaviors at different cooling rates such as the glass transition, melting point, crystallization, glass forming ability, and physical stability. In addition, broadband dielectric spectroscopy (BDS) has been widely used to determine the temperature dependence of relaxation times. The experimental timescale spans from $10^5$ ps to 100 s \cite{24,25,11,29,30,31,32,33}, which is far beyond simulation timescales. Thus, simulations cannot quantitatively predict temperature-dependent patterns and molecular-level features in experiments. From a theoretical point of view, one can employ the Elastically Collective Nonlinear Langevin Equation (ECNLE) theory to describe experiments quantitatively \cite{11,29,30,31,32,33,39}. This theory can predict structural relaxation times of up to $10^4$ s for multicomponent drugs. 

Gambogic acid, isolated from the resin of \emph{Garcinia hanburyi}, has various medical applications. It is a candidate for anti-tumor agents due to its ability to inhibit lung, gastric and colorectal cancers \cite{34,35,36}. The Chinese Food and Drug Administration approved gambogic acid for a phase II clinical trial for cancer treatment \cite{34,35,36}. Furthermore, this pharmaceutical material is expected to have low toxicity and side effects for the human body. Recently, gambogic acid has been used as a noncompetitive ligand for drugs to facilitate transport across biological barriers and binding to the cell surface \cite{37,38}.

In this work, we investigate the molecular dynamics and thermodynamic behaviors of gambogic acid in supercooled and glassy states using both experiments and ECNLE theory. These glassy properties of gambogic acid have not been studied before. We isolate gambogic acid from \emph{Garcinia hanburyi} trees planted in Vietnam. Then, these experimental samples are analyzed by BDS and DSC techniques. To gain better insight into the relaxation processes, we carry out theoretical calculations to describe experiments. The agreement between theory and experiment allows us to and understand nature of three different molecular relaxation dynamics and predict the effects of heating rate on the glassy dynamics of gambogic acid.
\section{Experimental Section}

\subsection{Plant Materials}
The resin of G. hanburyi was collected from Phu Quoc Island - Kien Giang Province, in December 2015. The plant material was identified by Dr. Nguyen Quoc Binh, Vietnam National Museum of Nature. The herbarium specimen has been deposited at the Institute of Natural Products Chemistry - Vietnam Academy of Science and Technology with the plant specimen number GH2015130.
\subsection{Extraction and Isolation}
The resin of \emph{Garcinia hanburyi} (500 g) was collected as a pale yellow aqueous suspension. The sample was diluted with acetone, and then was concentrated in vacuo to remove water. The dried resin (356 g) was extracted with methanol (MeOH) (3 L $\times$ 3) at room temperature using a conventional ultrasound-assisted technique. The methanol solution was condensed under reduced pressure to give an orange brown residue (257.0 g). The residue was further dissolved in dichloromethane(DCM) (500 mL $\times$ 3) to afford the DCM extract (89.0 g). The residue left behind was then extracted with ethylacetate (EtOAc) (500 mL $\times$ 3) to achieve the EtOAc extract (97.6 g).

The crude DCM extract was fractionated by column chromatography (CC) over silica gel, eluted with a gradient of n-hexane-EtOAc (v/v, 100:0 to 3:1), a gradient of DCM-EtOAc (v/v, 15:1 to 3:1) and a gradient of DCM-MeOH (v/v, 9:1 to 1:2), respectively to yield twelve fractions (Frs. GHN1-GHN12). Fraction GHN6 (29.8 g), fraction GHN7 (5.4 g) and fraction GHN8 (10.8 g)  were loaded separately onto CC over silica gel eluted with a gradient of n-hexane-EtOAc (v/v, 20:1 to 0:100) to afford the subfractions GHN6.1-GHN6.6, GHN7.1-GHN7.5 and GHN8.1-GHN8.6. Then, all subfractions were spotted onto thin layer chromatography (TLC) plates using standard Sigma-Aldrich gambogic acid 95 $\%$ purity as a reference. After developing the TLC plates with appropriate solvents, the plates were observed under a UV lamp at 254 and 365 nm, and then were stained with vanilin-$\ce{H_2SO_4}$ 10 $\%$ solution.The subfractions containing gambogic acid as a major component were collected for further isolation. Impure gambogic acid was subjected repeatedly to CC over RP-18 silica gel, eluting with MeOH-$\ce{H_2O}$ (v/v, 5:1) followed by purification on CC over silica gel with an eluent of n-hexane-EtOAc (v/v, 20:1) to obtain 0.744 g of gambogic acid as an orange solid.

\subsection{Differential Scanning Calorimetry}
Thermodynamic properties of gambogic acid were examined using a Mettler-Toledo Differential Scanning Calorimetry (DSC) 1 STARe System. The measuring device was equipped with an HSS8 ceramic sensor having 120 thermocouples and a liquid nitrogen cooling accessory. The instrument was calibrated for temperature and enthalpy using indium and zinc standards. The sample was examined in an aluminum crucible (40 $\mu L$). All measurements were carried out in a temperature range from 273 $K$ to 373 $K$ with a heating rate equal to 10 $K/$min.

\subsection{Broadband Dielectric Spectroscopy}
We measured the dielectric loss spectra for gambogic acid using Novo-Control GMBH Alpha dielectric spectrometer. The frequency and temperature range for the experiments spanned from $10^{-1}$ to $10^6$ Hz, and from 153 $K$ to 411 $K$, respectively. We used a Quattro temperature controller to manipulate the heating process with a thermal stability better than 0.1 $K$. The samples were examined in a parallel-plate cell made of stainless steel (diameter of 15 mm and a gap of 0.1 mm with glassy spacers). 

\section{Theoretical Approach}
\begin{figure}[htp]
\center
\includegraphics[width=9.5cm]{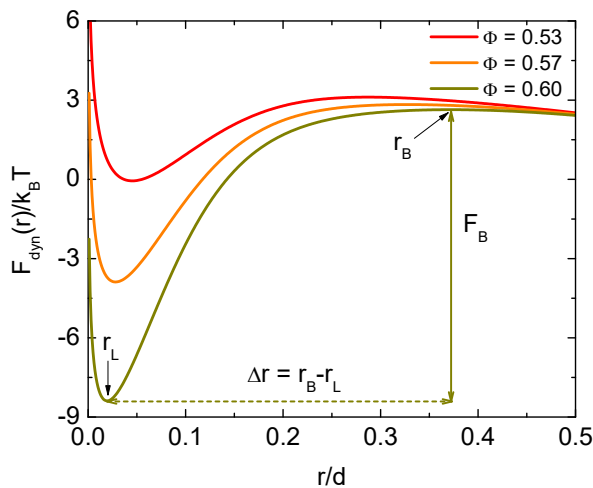}
\caption{(Color online) Dynamic free energy as a function of particle displacement for different packing fractions $\Phi=0.53$, 0.57, and 0.60. Important length and energy scales for the local dynamics are defined. }
\label{fig:5}
\end{figure}

The technical details of the ECNLE theory for amorphous drugs have been discussed in our previous work \cite{11,39}. Here, we briefly summarize the physical pictures. An amorphous material is viewed as a hard-sphere fluid of diameter $d$ and particle density $\rho$. When the density is sufficiently large, one observes a dynamical arrest of an arbitrary tagged particle within the nearest-neighbor length scale, $r_{cage}\approx 1.5d$, which is determined by the pair correlation function $g(r)$. The single molecule activated relaxation at temperature $T$ is analytically quantified using the dynamic free energy
\begin{eqnarray}
\frac{F_{dyn}(r)}{k_BT} &=& \int_0^{\infty} dq\frac{ q^2d^3 \left[S(q)-1\right]^2}{12\pi\Phi\left[1+S(q)\right]}\exp\left[-\frac{q^2r^2(S(q)+1)}{6S(q)}\right]
\nonumber\\ &-&3\ln\frac{r}{d},
\label{eq:5}
\end{eqnarray}
where $\Phi=\rho\pi d^3/6$ is a packing fraction, $S(q)$ is the static structure factor, $q$ is the wavevector, $k_B$ is the Boltzmann constant, and $r$ is the displacement of the tagged particle. The first term is known as a trapping potential caused by the nearest-neighbor interactions with the surrounding particles. The second term corresponds to the ideal fluid energy. As shown in Fig. \ref{fig:5}, the dynamic free energy profile gives a local barrier height, $F_B$, a localization length, $r_L$, a barrier position, $r_B$, and a jump distance, $\Delta r = r_B-r_L$. When the density is increased, motion of the tagged particle has more constraint. Thus, the jump distance is extended and the local barrier energy is raised with increasing $\Phi$.

The diffusion of the tagged particle from the particle cage requires a spatial reorganization of the other particles. The particle rearrangement generates an extra space at the cage surface and propagates a displacement field, $u(r)$, outside the cage via collective elastic fluctuation. The distortion field is proportional to $1/r^2$ and the amplitude is calculated by the jump distance. The physical treatment leads to a collective elastic barrier
\begin{eqnarray} 
F_{e} = 4\pi\rho\int_{r_{cage}}^{\infty}dr r^2 g(r)K_0\frac{u^2(r)}{2},
\label{eq:6}
\end{eqnarray}
where $K_0$ is the curvature of $F_{dyn}(r)$ at $r=r_L$, which determines the harmonic spring constant. Then, one can employ Kramer's theory to compute the structural or alpha relaxation time for a particle to escape from its particle cage 
\begin{eqnarray}
\frac{\tau_\alpha}{\tau_s} = 1+ \frac{2\pi}{\sqrt{K_0K_B}}\frac{k_BT}{d^2} e^{(F_B+F_e)/k_BT},
\label{eq:7}
\end{eqnarray}
where $K_B$ is the absolute curvature at $r = r_B$ and $\tau_s$ is a short relaxation time scale. The explicit expression for $\tau_s$ is given in our prior work \cite{11}. To determine the thermal response of the structural relaxation time and compare with experiments, an effective volume fraction of the hard-sphere fluids in the ECNLE calculations is analytically mapped to temperature via a physical picture of thermal expansion \cite{11}. The thermal mapping is
\begin{eqnarray}
T \approx T_0 - \frac{\Phi - \Phi_0}{\beta_l\Phi_0},
\label{eq:mapping1}
\end{eqnarray}
where $\Phi_0=0.5$ is the characteristic volume fraction and $\beta_l \approx 12 \times 10^{-4}$ $K^{-1}$ is a common value for the volume thermal expansion coefficient of organic materials in the supercooled-liquid state \cite{11}. Recall that the glass transition temperature $T_g$ measured by BDS and DSC techniques is defined by $\tau_\alpha(T=T_g) = 100$ s. Thus, a specific-material parameter $T_0$ is determined using the experimental $T_g$ and the ECNLE calculation $\tau_\alpha(\Phi\approx0.611) = 100$ s.

\section{Results and Discussion}
\subsection{Thermal Properties of Gambogic Acid}
\begin{figure}[htp]
\center
\includegraphics[width=8.5cm]{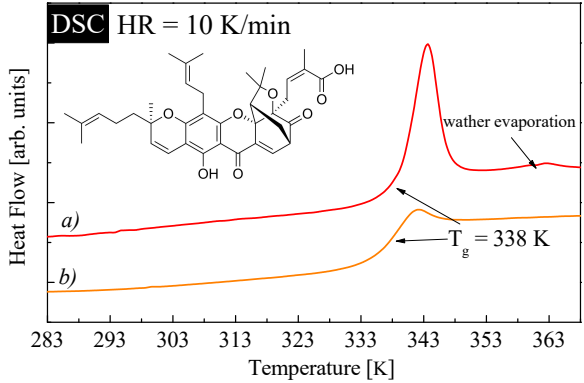}
\caption{(Color online) DSC thermograms for the (a) as-obtained gambogic acid and (b) gambogic acid annealed at 373 $K$ for 3 min prior to measurement. The inset presents the chemical structure of the investigated material.}
\label{fig:1}
\end{figure}

Thermal properties of gambogic acid were investigated by means of DSC. Figure \ref{fig:1} shows the thermograms for two runs: (i) measurement of the as-received sample (aged) and (ii) measurement of the sample annealed at 373 $K$ (rejuvenated) prior to the DSC experiment. Both samples were measured by heating from 273 $K$ to 373 $K$ with a rate equal to 10 $K/$min. From the thermogram of the aged sample, one can observe an endothermic peak superimposed onto the endothermic step of the experimental heating curve associated with the glass-to-supercooled liquid transition. This clearly visible thermal process, reflecting a regaining of the entropy/enthalpy lost during the phase transition, indicates that the measured glassy material was indeed stored in the amorphous form for a longer period of time \cite{1}. During further heating for the DSC thermogram of this sample, a second endothermic peak having much lower amplitude was registered. Since this broad peak is located in the vicinity of 353-373 $K$, it can be associated with the water evaporation. To remove the water, prior to all further experiments, the sample was annealed for 3 min at 373 $K$. The DSC curve for the annealed, and at the same time rejuvenated, sample exhibits a step-like thermal event with an overlapping endothermal peak with a much lower $\Delta H$ than in the case of the aged sample. Additionally, the DSC trace did not reveal the presence of the peak associated with water evaporation. The glass transition of neat gambogic acid was determined as the midpoint of the heat capacity increases, which occurs at a temperature equal to 338 $K$.

\subsection{Molecular Dynamics of Supercooled and Glassy Gambogic Acid}
\begin{figure}[htp]
\center
\includegraphics[width=9cm]{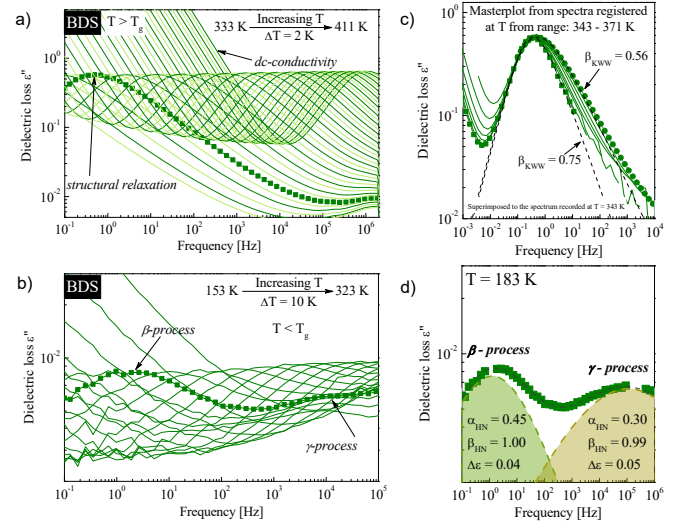}
\caption{(Color online) Dielectric loss spectra for gambogic acid obtained upon heating. The spectra collected (a) above and (b) below the sample's glass transition temperature. (c) The masterplot constructed by superimposing dielectric spectra obtained at eight different temperatures above the glass transition temperature. The dashed black lines represent the KWW fit. (d) An  example of the fitting procedure used for the spectrum registered at $T = 183$ $K$.}
\label{fig:2}
\end{figure}

We measure the dielectric loss spectra in the broad frequency range from $10^{-1}$ Hz to $10^6$ Hz to investigate the molecular mobility of amorphous gambogic acid. During this measurement, the temperature increases from 153 $K$ to 333 $K$ with a step size of 10 $K$, and from 333 $K$ to 411 $K$ with a step size of 2 $K$. The obtained dielectric loss spectra registered both in the supercooled and glassy states are presented in Figure \ref{fig:2}a and b, respectively. In the temperature region below the drug's glass transition temperature, two secondary relaxation processes -$\beta$ and $\gamma$- associated with the local (inter- or intramolecular) motions are noticeable in the dielectric loss spectra for gambogic acid \cite{2}. While the dielectric loss spectra registered at $T > T_g$ exhibit one well-resolved loss peak corresponding to the structural $- \alpha -$ relaxation as well as dc-conductivity. All relaxation processes for gambogic acid move toward higher frequencies with increasing temperature, indicating an increase in molecular mobility.

The thermal response of the $\alpha$-peak can be analyzed using a masterplot. The dielectric spectra taken at temperatures in the region of 343-371 $K$ are horizontally shifted and superimposed onto the reference spectrum at 343 $K$. The masterplot shown in Figure \ref{fig:2}c clearly indicates that the shape of the peak of structural relaxation for gambogic acid becomes narrower with increasing temperature. Then, we employ one-side Fourier transformation of the Kohlrausch-Williams-Watts (KWW) function \cite{3,4} to fit the $\alpha$-loss peak and determine the parameter $\beta_{KWW}$. This parameter describes the breadth of the $\alpha$-relaxation peak and varies between 0.56 and 0.75.

Next, we determine the relaxation time for the $\alpha-$, $\beta-$ and $\gamma-$ processes. To obtain the values of $\tau_\alpha$, $\tau_\beta$, and $\tau_\gamma$, the asymmetric structural relaxation process and symmetric secondary relaxation process are fitted using Havriliak-Negami (HN) and Cole-Cole (CC) functions, respectively. The empirical HN function is \cite{5}: 
\begin{eqnarray}
\varepsilon^*_{HN}(\omega)=\varepsilon' (\omega) - i\varepsilon'' (\omega)=\varepsilon_\infty + \frac{\Delta \varepsilon}{\left[1+(i\omega\tau_{HN})^a \right]^b},
\label{eq:1}
\end{eqnarray}
where $\varepsilon'(\omega)$ and $\varepsilon''(\omega)$ are the real and imaginary parts of the complex dielectric function, respectively, $\omega$ is equal to $2\pi f$, $\Delta \varepsilon$ is the dielectric strength, and $\varepsilon_\infty$ is the high frequency limit permittivity, and $\tau_{HN}$ is the HN relaxation time. The parameter $a$ and $b$ represent the symmetric and asymmetric broadening of the relaxation peak, respectively. It has to be noted that when the parameter $b$ is equal to unity, the HN function becomes the CC function to fit the secondary relaxations of gambogic acid. An example of the fitting procedure performed for the spectrum registered at 183 $K$, is presented in Figure \ref{fig:2}d. Based on these fitting parameters, one can calculate $\tau_\alpha$, $\tau_\beta$, and $\tau_\gamma$ by employing the following equation:
\begin{eqnarray}
\tau_\alpha = \tau_{HN}\left[\sin\left(\frac{\pi a}{2+2b} \right) \right]^{-1/a}\left[\sin\left(\frac{\pi ab}{2+2b} \right) \right]^{1/a}.
\label{eq:2}
\end{eqnarray}

\begin{figure}[htp]
\center
\includegraphics[width=9cm]{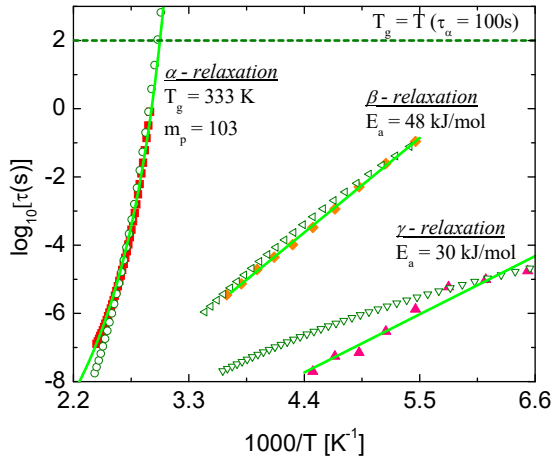}
\caption{(Color online) Relaxation map of gambogic acid. Temperature dependence of $\tau_\alpha$ in the supercooled liquid described by the VFT equation. The temperature dependence of $\tau_\beta$ and $\tau_\gamma$ was fitted using the Arrhenius equation. The open data points and solid curves correspond to our ECNLE calculations and the fit function, respectively. The solid data points are experimental results.}
\label{fig:3}
\end{figure}

Figure \ref{fig:3} shows the temperature dependence of the $\alpha-$, $\beta-$ and $\gamma-$ relaxation times for gambogic acid. Our ECNLE calculations for $\tau_\alpha(T)$ agree quantitatively well with the experimental data counterpart measured by BDS. We also describe the dielectric data for the $\alpha-$ process using the Vogel-Fulcher-Tammann (VFT) equation \cite{6,7,8}
\begin{eqnarray}
\tau_\alpha = \tau_\infty\exp\left(\frac{B}{T-T_{VFT}} \right),
\label{eq:3}
\end{eqnarray}
where the fitting parameters correspond to $\log\tau_\infty = -13.1 \pm 0.2$, $T_{VFT} = 265.3 \pm 2.0$, and $B=2100.9 \pm 87$. The overlap for the theoretical curve, experimental data points, and the VFT curve is quite good. This finding validates the ECNLE theory for predicting the glassy dynamics of amorphous materials without any adjustable/free parameter. By extrapolating the VFT equation to $\tau_\alpha = 100$ s, we estimate the glass transition temperature of the examined material to be equal to 333 $K$, which is 5 $K$ lower than that found in the DSC study at a heating rate of 10 $K/min$. This discrepancy is somewhat high but not unusual.

We can use $\tau_\alpha$ at room temperature to estimate the physical stability time or how the glassy gambogic acid is stable during storage at room temperature \cite{43}. From our the theoretical temperature dependence of structural relaxation time, $\tau_\alpha(T=300K)\approx 10^{14.3}$s. The numerical result indicates that the investigated material may be stable after approximately $2.31\times 10^9$ days. Beyond the time scale, some crystallites may occur when stored at room temperature.

Figure \ref{fig:3} also shows that gambogic acid undergoes two secondary relaxations ($\beta-$ and $\gamma-$ process). These relaxation times exhibit the usual Arrhenius temperature dependence. The values for the activation energy $E_a$ of the secondary relaxation processes of gambogic acid can be determined by fitting the $\tau_\beta(T)$ and $\tau_\gamma(T)$ dependence with the Arrhenius equation. Our numerical results reveal that the activation energy of the $\beta$-process ($\sim$ 48 $kJ/mol$) is relatively larger than that of the $\gamma-$process ($\sim$ 30 $kJ/mol$). This suggests that the molecular dynamics in the $\gamma$-process is faster.

There are two main interpretations for origin of the secondary relaxations: (i) intramolecular motions as they are far away from the alpha relaxation, and (ii) the single-molecule relaxation known as Johari-Goldstein relaxation. The latter type of process in our ECNLE theory can be viewed as local dynamics of a single particle within its particle cage that is unrelated to the molecular rearrangement beyond the first shell. Thus, only the local barrier $F_B$ of the dynamic free energy $F_{dyn}(r)$ contributes to the Johari-Goldstein secondary relaxation. The Johari-Goldstein relaxation time now is
\begin{eqnarray}
\frac{\tau_{JG}}{\tau_s} = 1+ \frac{2\pi}{\sqrt{K_0K_B}}\frac{k_BT}{d^2} e^{F_B/k_BT}.
\label{eq:11}
\end{eqnarray}

The thermal expansion coefficient used in the thermal mapping for the alpha relaxation is different from that for the secondary relaxation. Without particle escape, the structure of amorphous materials in the Johari-Goldstein relaxation seems to be "frozen". The vibrational motion of the tagged particle is similar to a phonon-like mode. Since crystal-like structures expand thermally harder than the disordered counterparts, the thermal expansion coefficient in Eq. (\ref{eq:mapping1}) for the secondary relaxation should be reduced. To gain the best agreement with our experimental data of the $\beta-$ and $\gamma-$process, we choose $\beta_l\rightarrow \beta_g = 8.4 \times 10^{-4}$ and $6 \times 10^{-4}$ $K^{-1}$, respectively. As shown in Fig. \ref{fig:3}, our ECNLE calculations agree quantitatively well with the $\beta$ relaxation time while the deviation between theory and experiment in the $\gamma$ relaxation is relatively significant. These findings clearly indicate the strong relation between the $\beta-$ and $\alpha-$relaxation. Particularly, the $\beta-$process is the precursor of the $\alpha-$process and this interpretation is consistent with other works \cite{40,41}. Our calculations also possibly suggest a new approach to estimate the thermal expansion coefficient in glassy state. The proposal need further experiments to test. In contrast, the $\gamma$ relaxation must be due to an intramolecular process.

The effects of heating rate, $h$, on the glass transition temperature can be investigated using the ECNLE theory. Near $T_g$, it is possible to define the heating rate as \cite{12}
\begin{eqnarray}
h = \frac{dT}{dt}=-\frac{dT}{d\tau_\alpha}.
\label{eq:8}
\end{eqnarray}

After taking the natural logarithm of both sides of Eq. (\ref{eq:7}) and associating with Eq. (\ref{eq:8}), one obtains an approximate expression to correlate $h$ with $\tau_\alpha(T_g)$ and two dynamic barriers in the ECNLE theory at $T = T_g$
\begin{eqnarray}
\left.h\tau_\alpha(T_g)\frac{d}{dT}\left( \frac{F_B+F_e}{k_BT}\right)\right|_{T=T_g} \approx -1.
\label{eq:9}
\end{eqnarray}

\begin{figure}[htp]
\center
\includegraphics[width=8.5cm]{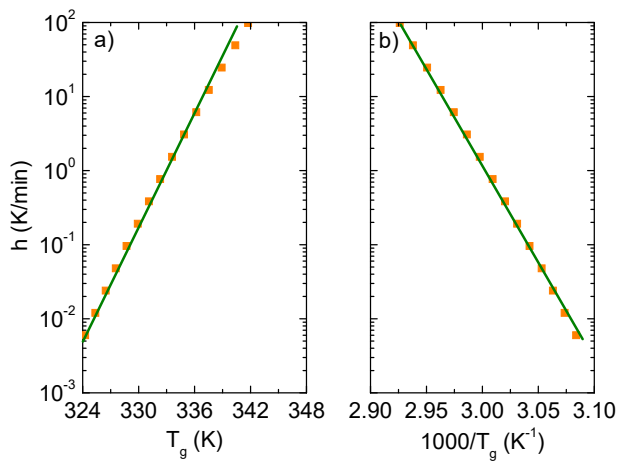}
\caption{(Color online) The heating rate as a function of (a) $T_g$ and (b) $1000/T_g$ for gambogic acid. The data points show our theoretical predictions and the solid lines are a guide to the eye.}
\label{fig:4}
\end{figure}

Figure \ref{fig:4} shows the heating rate dependence of $T_g$ calculated using the ECNLE theory. Interestingly, the heating rate exponentially varies with both $T_g$ and $1000/T_g$. The Arrhenius form of $h$ is consistent with many studies \cite{15,16,17,18,20,22}. The linear relationship between $\log_{10}h$ and $T_g$ can also be found in Ref. \cite{19,10,21}. Interestingly, the theoretical results clearly indicate that $T_g = 338 K$ at $h \approx 10 K/min$. This value quantitatively agree with the DSC result in this work. As seen in Figure \ref{fig:3}, when $\tau_\alpha$ ranges from $10^{-4}$ ps to 1 s, the temperature dependence of $\tau_\alpha$ seems to obey Arrhenius behavior. This suggests that the total barrier, $F_B+F_e$, is nearly constant at low temperatures. One can simplify Eq. (\ref{eq:9}) to be
\begin{eqnarray}
\frac{h\tau_\alpha(T_g)}{T_g^2}   = \ce{constant}.
\label{eq:10}
\end{eqnarray}
This expression is similar to that obtained in the previous results \cite{13,14}.

One can use the VFT extrapolation to calculate the fragility parameter ($m_p$), which is defined as
\begin{eqnarray}
m_p = \left.\frac{d\log_{10}\tau_\alpha}{d\left(T_g/T \right)}\right|_{T=T_g}.
\label{eq:4}
\end{eqnarray}

The typical values for $m_p$ for various materials vary between 16 and 200 \cite{9}. According to the Angell approach, the fragility parameter is typically used to classify supercooled liquids into three categories: \emph{fragile}, \emph{intermediate} or \emph{strong}. So-called \emph{strong} liquids are characterized by small values of this parameter (ca. 16-60), while large values of $m_p$ indicate that the liquid is fragile and that its $\tau_\alpha(T)$ is more Arrhenius-like. Thus, gambogic acid with $m_p$ equal to 103, can be classified as a fragile liquid. Phan and his coworkers \cite{42} used ECNLE theory and values of $T_g$ and $m_p$ to understand the correlation between local and cooperative motions of molecules. Low-fragility materials have a low elastic barrier or low cooperative character compared to the local barrier or local dynamics. Similarly, effects of collective motions in high-fragility materials on the glass transition are stronger than those of local motions. Moreover, in our above calculations, we clearly show that $T_g$ is remarkably dependent on the heating rate. One can expect that a strong dependence of the fragility on the heating rate. 

\section{Conclusions}
We have employed experiments and ECNLE theory to study the relaxation dynamics for gambogic acid at the microscopic level. The glass transition temperature ($T_g \approx 338$ $K$) representing the molecular mobility in the amorphous systems is measured using DSC measurement at a heating rate of 10 $K/min$. Then we have used BDS measurements to determine the thermal response of the $\alpha-$, $\beta-$, and $\gamma-$relaxation times. By fitting our BDS experimental data for $\tau_\alpha(T)$ with the VFT equation, we obtain $T_g \approx 333$ $K$ and a dynamic fragility parameter of $m_p \approx 103$. This result slightly differs from the DSC measurement but the deviation is acceptable and usual. To understand deeply these experimental results, we have applied the ECNLE theory to calculate the temperature dependence of the structural and secondary relaxation times and glass transition temperature for a wide range of heating rates. Our theoretical calculations for $\tau_\alpha(T)$ are similar to those found in the experiments without any adjustable parameter. In addition, we clearly reveal that the $\beta-$process has intrinsic and strong correlation with the structural relaxation, and arise from the local dynamics of a single molecule. The relaxation is known as Johari-Goldstein relaxation. Meanwhile, the $\gamma-$process possibly originates from intramolecular motions. The predicted $T_g$ as a function of heating rate agrees quantitatively well with that found in our DSC and BDS experiments. The logarithm of the heating rate is linearly proportional to $T_g$ and $1000/T_g$. This behavior is qualitatively consistent with previous studies.

\begin{acknowledgments}
This work was supported by JSPS KAKENHI Grant Numbers JP19F18322 and JP18H01154. The author, M.P., is grateful for the financial support received within the Project No. 2015/16/W/NZ7/00404 (SYMFONIA 3) from the National Science Centre, Poland. This work was also supported by Ministry of Industry and Trade (Vietnam), project CNHD 061/15-17. This research was funded by the Vietnam National Foundation for Science and Technology Development (NAFOSTED) under grant number 103.01-2019.318.
\end{acknowledgments}

\end{document}